\documentclass[12pt]{iopart}

\usepackage{epsfig}
\usepackage{graphicx}
\usepackage{lscape}
\begin{document}

\title[]{Recognition of the First Observational Evidence of an Extrasolar Planetary System}

\author{B. Zuckerman$^1$}

\address{$^1$Department of Physics and Astronomy, University of California, Los Angeles, CA 90095, USA}
\eads{\mailto{ben@astro.ucla.edu}}
\begin{abstract}
With 20-20 hindsight, it is now possible to say that the $\it{first}$ observational indication -- by any means -- of the existence of an extrasolar planetary system came almost a century ago when van Maanen discovered and noted the spectrum of the nearest single white dwarf to Earth.  
\end{abstract}

\section{Introduction}

In late 2013, Dr. Jay Farihi invited me to give a review talk on the compositions of extrasolar planetesimals and rocky planets at a July 2014 symposium on "Characterizing Planetary Systems Across  the HR Diagram".    He suggested that I include in my review a brief history of the field of heavy element "pollution" of the photospheres of white dwarfs.   While preparing this history I experienced a true "eureka" moment when I realized that the very first observational evidence for the existence of an extrasolar planetary system came nearly 100 years ago from one of these polluted stars, van Maanen 2.  However, proper interpretation of this evidence in terms of the existence of a planetary system in orbit around van Maanen 2 did not come for about another 90 years.  The present discussion briefly outlines the story.

\section{Discussion}

A decade of observational and theoretical studies by many astronomers has shown that the presence of heavy elements in the spectrum of a single white dwarf star with effective temperature in the range 5000 to 20,000 K is evidence for the presence of a wide-orbit planetary system (see references listed in the Introduction in Zuckerman 2014).  In most, perhaps all, cases the planetary system contains at least one rocky debris belt and one major planet (e.g., Veras et al 2013; Mustill et al 2014).   The gravitational field of the planet(s) can perturb the orbits of asteroids so that, following a sequence of events, rocky asteroidal material is eventually accreted onto the white dwarf (e.g. Jura \& Young 2014).  To the best of my knowledge, no other model for the heavy element pollution of white dwarf atmospheres in this temperature range is now seriously considered.

The first discovered white dwarf with heavy elements in its photosphere was van Maanen 2 (van Maanen 1917 \& 1919) although at the time van Maanen did not know that he had observed a white dwarf, nor were the nature of white dwarfs even understood.   A modern spectrum of van Maanen 2 appears in Figure 1.   The very strong metal lines in the low resolution spectra obtained 100 years ago led 
van Maanen to classify the star as F-type while noting its very low bolometric luminosity for this spectral class, "...by far the faintest F-type star known at the present time" (van Maanen 1919).

As time went on more white dwarfs with photospheric metals were discovered, while at the same time it was appreciated that the intense gravitational fields of such stars should cause these heavy elements to sink rapidly out of sight (Schatzman 1945).  Thus a source of the metals was required; for decades the favored source was accretion of material from the interstellar medium.   But a series of observational papers, a few noteworthy ones are listed in Table 1, accompanied by a series of interpretive papers -- early ones include Graham et al 1990, Debes \& Sigurdsson 2002, and Jura 2003 -- gradually shifted the accretion paradigm from interstellar material to rocky asteroidal debris.  If one dates the triumph of the latter model over the former to about 2007, then there was a delay of about 90 years between the discovery of heavy elements in the spectrum of van Maanen 2 and their proper interpretation as originating in a surrounding planetary system.

\section*{REFERENCES}
\begin{harvard}

\item[Debes, J. \& Sigurdsson, S. 2002, ApJ 572, 556] 
\item[G{\"a}nsicke, B., Marsh, T., Southworth, J. \& Rebassa-Mansergas, A. 2006, Science 314, 1908] 
\item[Graham, J., Matthews, K., Neugebauer, G. \& Soifer, B. T. 1990, ApJ 357, 216] 
\item[Jura, M. 2003, ApJ 584, L91] 
\item[Jura, M. \& Young, E. 2014, Ann. Rev Earth Planet Sci, 42, 45] 
\item[Mustill, A., Veras, D. \& Villaver, E. 2014, MNRAS 437, 1404] 
\item[Napiwotzki, R., Christlieb, N., Drechsel, H. et al. 2003, Msngr, 112, 25] 
\item[van Maanen, A. 1917, PASP 29, 258] 
\item[van Maanen, A. 1919, AJ 32, 86] 
\item[Schatzman, E. 1945, Ann. Astrophys. 8, 145] 
\item[Veras, D., Mustill, A., Bonsor, A. \& Wyatt, M. 2013 MNRAS 431, 1686] 
\item[Zuckerman, B. 2014, ApJ 791, L27] 
\item[Zuckerman, B. \& Becklin, E. E. 1987, Nature 330, 138] 
\item[Zuckerman, B., Koester, D., Melis, C., Hansen, B. \& Jura, M. 2007, ApJ 671, 872] 
\item[Zuckerman, B., Koester, D., Reid, I. N. \& Hunsch, M. 2003, ApJ 596, 477] 

\end{harvard}

\clearpage

\begin{table}
\caption{Some Discovery Papers on Accretion by White Dwarfs of Planetary Debris}
\begin{tabular}{@{}lcc}
\br
Year &  Discovery & Reference \\
\mr
1917 &  metals in a white dwarf atmosphere  &  van Maanen 1917 \\
1987 &  a hot dusty disk is orbiting a white dwarf  &   Zuckerman \& Becklin 1987 \\
2003 &  25\% of H-atmosphere white dwarfs contain metals &  Zuckerman et al 2003 \\
2006 &  a gaseous disk is orbiting a white dwarf  &  G{\"a}nsicke et al 2006 \\  
2007 & a white dwarf atmosphere exhibiting 17 elements & Zuckerman et al 2007 \\
\br
\end{tabular}
\end{table}

%
%

\begin{figure}
\includegraphics[width=180mm]{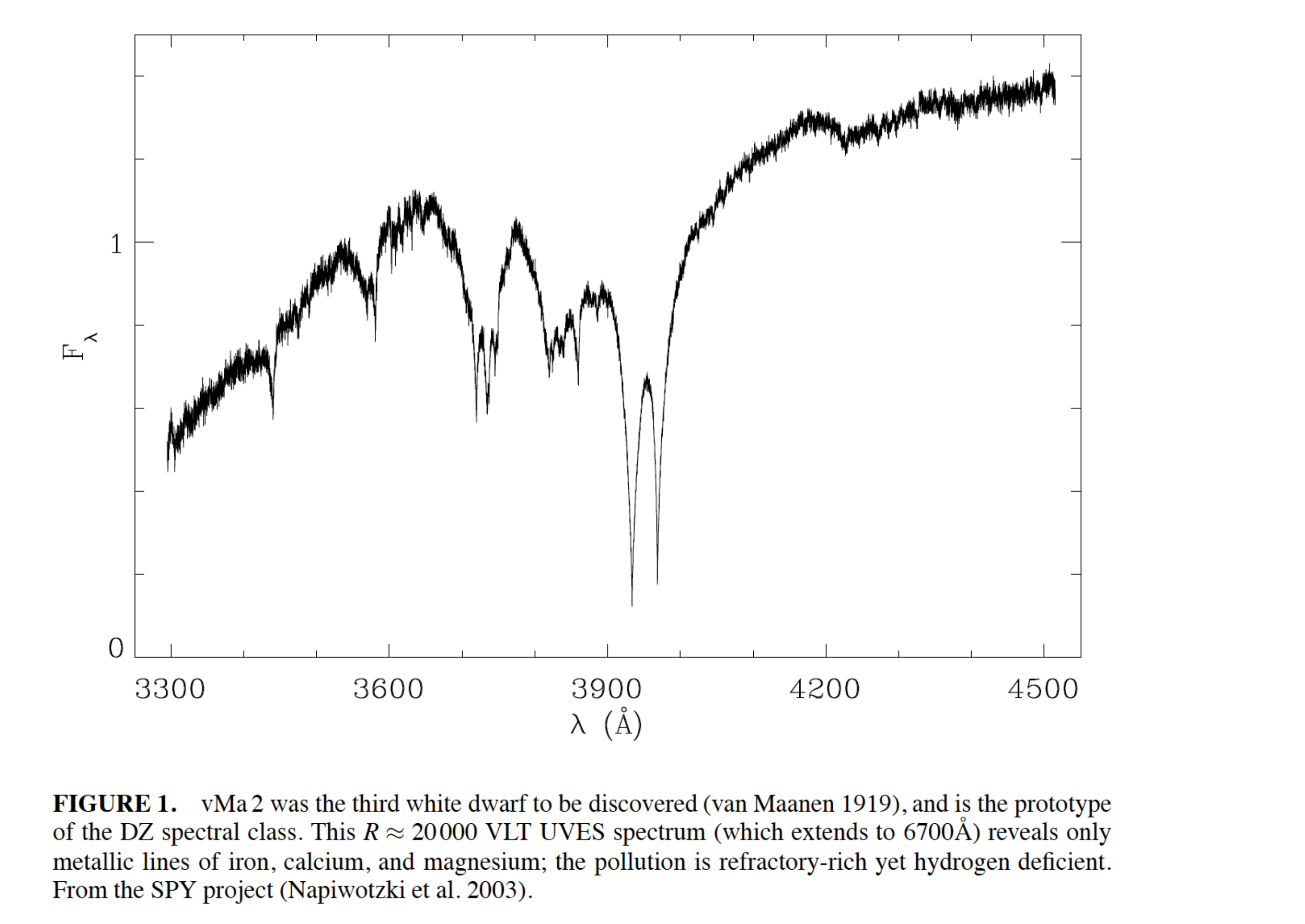}
\end{figure}

\end{document}